\documentclass[10pt,preprint,a4paper]{article}

\usepackage{amsmath, amssymb, mathrsfs, graphicx, subcaption, dsfont, enumerate, epstopdf, xcolor}
\usepackage{cite} 
\usepackage[colorlinks, allcolors=blue!70!black, linktocpage]{hyperref}
\usepackage[inline]{enumitem}
\usepackage[utf8]{inputenc}

\usepackage{epic}
\usepackage{eepic}
\usepackage{graphicx}

\numberwithin{equation}{section}

\begin{document}
\begin{titlepage}

$~$
\vskip 4.0cm

\begin{center}
{\Large \bf Comments on the multi-spin solution to the Yang-Baxter equation  \\ \vspace{0.2cm}  and   basic hypergeometric sum/integral identity}
\end{center}
\vskip 1.5 cm

\centerline{\large {\bf Ilmar Gahramanov\footnote{ilmar.gahramanov@msgsu.edu.tr}$\,^{a,b,c,d}$} and
{\bf Shahriyar Jafarzade\footnote{shahriyar.jzade@gmail.com}$\,^{a,b,c}$} }

\begin{center}

\textit{$^{a}$ Department of Physics, Mimar Sinan Fine Arts University,\\ Bomonti 34380, Istanbul, Turkey} \\
\texttt{} \\
\vspace{.2mm}
\vspace{.2mm}
\textit{$^{b}$ Department of Mathematics, Khazar University, \\ Mehseti St. 41, AZ1096, Baku, Azerbaijan} \\
\texttt{} \\
\vspace{.2mm}
\textit{$^{c}$ Institute of Radiation Problems ANAS,\\ B.Vahabzade 9, AZ1143 Baku, Azerbaijan} \\
\texttt{} \\
\textit{$^{d}$  Max Planck Institute for Gravitational Physics (Albert Einstein Institute),\\ Am M\"{u}hlenberg 1, D-14476 Potsdam, Germany} \\
\texttt{}

\vspace{0mm}
\end{center}

\vskip 1.5cm 
\begin{center}
{\bf Abstract} \vskip 0.2cm
\end{center}
\noindent We present a multi-spin  solution to  the Yang-Baxter equation. The solution corresponds to the integrable lattice spin model of statistical mechanics with positive Boltzmann weights and  parameterized in terms of  the basic hypergeometric functions. We obtain this solution from a non-trivial basic hypergeometric sum-integral identity which originates from the equality of supersymmetric indices for  certain three-dimensional $\mathcal N=2$ Seiberg dual theories.

\end{titlepage}





\section{Introduction and Conclusions}

There has recently been a lot of interest in the study of integrability phenomenon in, and relation to quantum field theory (see, e.g. \cite{Morozov:2013xra,Lamers:2015dfa,Negro:2016yuu,Fadeev}), especially to supersymmetric gauge theories. The integrability via  the Yang-Baxter equation  turns out to be of particular interest since its appearance in several areas of mathematical physics, particularly it plays a fundamental role in integrable models of statistical mechanics (see, e.g., \cite{baxter2007exactly,Jimbo:1989mc,Jimbo:1989qm,Bazhanov:2016ajm,Batchelor:2015osa}).

The search for solutions to the Yang-Baxter equation in the context of the so-called gauge/YBE correspondence is one of the great successes in the subject \cite{Spiridonov:2010em,Yamazaki:2012cp,Kels:2015bda,Yagi:2015lha,Gahramanov:2016ilb,Gahramanov:2015cva, Yamazaki:2015voa,Yamazaki:2013nra,Kels:2017toi,Jafarzade:2017fsc,Kels:2017vbc}. It is known that every solution to the Yang-Baxter equation corresponds to an integrable model of statistical mechanics on a two-dimensional lattice. The gauge/YBE correspondence establishes a relation between supersymmetric quiver gauge theories and two-dimensional lattice spin models.

In these notes we present a new solution to the Yang-Baxter equation in terms of basic hypergeometric functions which is a multi-spin
version of the exactly solvable model found in \cite{Gahramanov:2015cva,Kels:2015bda}. The Yang-Baxter equation for this model becomes a new non-trivial basic-hypergeometric sum/integral identity. We found the solution by identification the equality of supersymmetric indices for certain supersymmetric dual theories with the star-star relation for the integrable lattice models.

There exist many possible future directions and we list some of them. A new  basic hypergeometric sum/integral identity  conjectured in this note   needs to be proven (the univariate integrals were considered in \cite{Krattenthaler:2011da,Gahramanov:2013rda,Gahramanov:2014ona,Gahramanov:2016wxi,Rosengren:2016mnw} ). Another possible future direction is to study the aspects of this solution in the framework of quantum algebras along the lines of \cite{Chicherin:2014dya,Chicherin:2015mfv}.

This paper is organized as follows. We start by recalling the solution to the star-triangle relation (a special form of the Yang-Baxter equation) with discrete and continuous spin variables found in \cite{Gahramanov:2015cva,Kels:2015bda} in Section 2. A new basic hypergeometric sum/integral identity and the solution to the Yang-Baxter equation is presented in Section 3. In the last section, the relation between basic hypergeometric sum/integral identity and  three-dimensional supersymmetric index is briefly discussed.

\section{Ising-type models with discrete and continuous spin}

There exist two-dimensional lattice models in statistical mechanics which are called Ising-type or edge-interaction models where the spins are located on sites of the lattice and interacted only with nearest neighbors. If the Boltzmann weights of such model satisfies  the star-triangle relation, then the model is integrable. There are many solutions to the star-triangle relation,
some of them are found from supersymmetric gauge theory calculations\footnote{The complete list of references is given in our upcoming review paper\cite{Jafarzade-2017}.}  within a few years \cite{Spiridonov:2010em,Kels:2015bda,Gahramanov:2016ilb,Gahramanov:2015cva, Jafarzade:2017fsc}. The star-triangle relation which leads Ising-type models with discrete and continuous spin variables has the following form
\begin{align}\nonumber
     \sum_{m_0} \int d x_0\, \mathcal{S}( x_0|m_0 ) &\mathcal{W}_{\eta-\gamma}(x_0|m_0,x_i|m_i) \mathcal{W}_{\eta - \beta}( x _{j}|m_j, x_0|m_0) \mathcal{W}_{ \eta - \alpha}( x_{k}|m_k , x_0|m_0)\\ \label{star-triangle}
    & \qquad = \mathcal{W}_{ \alpha }( x _{i}|m_i, x_{j}|m_j)\mathcal{W}_{ \beta}( x _{k}|m_k, x_{i}|m_i)\mathcal{W}_{ \gamma}( x _{k}|m_k, x_{j}|m_j),
\end{align}
where $\eta=\alpha+\beta+\gamma$ is the crossing parameter, $\mathcal{W}$ is the Boltzmann weight describe the interaction between spin points , $\mathcal{S}$ is the self interaction term, $x_i$ and $m_i$ are continuous and discrete spin variables, respectively. 

There are a few known solutions to the star-triangle relation of the form of (\ref{star-triangle}), i.e. the model with discrete and continuous spin variables \cite{Kels:2013ola,Gahramanov:2015cva,Kels:2015bda,Gahramanov:2016ilb}.  We are interested
in the solution \cite{Gahramanov:2015cva,Kels:2015bda} originating from the three-dimensional $\mathcal N = 2$ supersymmetric
index calculations \cite{Krattenthaler:2011da,Gahramanov:2013rda,Gahramanov:2014ona,Gahramanov:2016wxi,Gahramanov:2015tta} which has the following form
\begin{align} \nonumber
        \mathcal{W}_{\alpha}(x_i|m_i,x_j|m_j) & =\Phi_\alpha(x, m) \frac{(q^{1+(m_i+m_j)/2}q^{\eta-\alpha-i(x_i+x_j)};q)_\infty}{(q^{(m_i+m_j)/2}q^{\alpha-\eta+i(x_i+x_j)};q)_\infty}   \frac{(q^{1+(m_j-m_i)/2}q^{\eta-\alpha+i(x_i-x_j)};q)_\infty}{(q^{(m_j-m_i)/2}q^{\alpha-\eta-i(x_i-x_j)};q)_\infty} \\ \label{W}
 & \quad \times   \frac{(q^{1-(m_i+m_j)/2}q^{\eta-\alpha+i(x_i+x_j)};q)_\infty}{(q^{-(m_i+m_j)/2}q^{\alpha-\eta-i(x_i+x_j)};q)_\infty}    \frac{(q^{1+(m_i-m_j)/2}q^{\eta-\alpha+i(x_j-x_i)};q)_\infty}{(q^{(m_i-m_j)/2}q^{\alpha-\eta-i(x_j-x_i)};q)_\infty}, \\ \label{S}
        \mathcal{S}(x_0|m_0) & = \frac{1}{q^{m_0}}\frac{(q^{2x_0+m_0};q)_\infty}{(q^{2x_0+m_0+1};q)_\infty}\frac{(q^{-2x_0+m_0};q)_\infty}{(q^{-2x_0+m_0+1};q)_\infty},
\end{align}
where
\begin{equation}\label{k}
\Phi(\alpha, x, m) \ = \ \frac{q^{-2i(x_im_i+x_jm_j)}} {k(\alpha)} ,\; \; \text{and} \;\; k(\alpha)=\exp\Big(-\sum_{n\neq 0}\frac{e^{4\alpha n}}{n(q^n-q^{-n})}\Big).
\end{equation}
The continuous and discrete spin variables for the model are $ 0 \leq x_j < 1$ and $ m_j \in Z$, respectively. Here we used the  usual q-Pochhammer symbol $(z;q)_{\infty}:=\prod_{i=0}^{\infty}(1-zq^i)$. 

This solution to the star-triangle relation (\ref{star-triangle}) can be obtained from the following basic hypergeometric sum/integral identity \cite{Gahramanov:2016wxi,Gahramanov:2015cva,Rosengren:2016mnw}
\begin{align}\label{str-tri}\nonumber
     \sum_{m=-\infty}^{\infty}  \oint \prod_{i=1}^6 \frac{(q^{1+(m+n_i)/2}/a_iz;q)_\infty(q^{1+(n_i-m)/2} z/a_i ;q)_\infty}{(q^{(m+n_i)/2}a_iz;q)_\infty(q^{(n_i-m)/2} a_i/z ;q)_\infty} \frac{(1-q^m z^2)(1-q^m z^{-2})}{q^m z^{6m}} \frac{dz}{2\pi i z} \\  =\frac {2}{ \prod_{i=1}^6 a_i^{n_i}}  \prod_{1 \leq i<j  \leq 6} \frac {(q^{1+(n_i+n_j)/2}/a_i a_j;q)_\infty}{(q^{(n_i+n_j)/2}a_i a_j;q)_\infty},
\end{align}
with the balancing conditions $\prod_{i=1}^6 a_i = q$ and $\sum_{i=1}^6 n_i =  0$. The proof of this identity  is presented in \cite{Gahramanov:2016wxi}.

Another type of the Yang-Baxter equation is the IRF-type. The solution to the IRF-type Yang-Baxter equation describes an integrable model in which  four spins are located around a face of the lattice, and  interact with each other. We are interested in the following form of the IRF-type Yang-Baxter equation  
\begin{align}\label{Irf-ybe}
\begin{array}{l}
\displaystyle \sum_{H\in Z} \int [d_H h]\;
\mathbb{R}_{(t_4,t_1)(t_6,t_3)}\left(\begin{array}{cc}
{a|A} & {b|B} \\
{h|H} & {c|C}\end{array}\right)
\mathbb{R}_{(t_6,t_3)(t_2,t_5)}\left(\begin{array}{cc}
{c|C} & {d|D} \\
{h|H} & {e|E}\end{array}\right)
\mathbb{R}_{(t_2,t_5)(t_4,t_1)}\left(\begin{array}{cc}
{e|E} & {f|F} \\
{h|H} & {a|A}\end{array}\right)\\
[5mm]
\displaystyle \;\;\;\;
=\sum_{H\in Z}\int [d_H h]\;
\mathbb{R}_{(t_6,t_3)(t_2,t_5)}\left(\begin{array}{cc}
{b|B} & {h|H} \\
{a|A} & {f|F}\end{array}\right)
\mathbb{R}_{(t_2,t_5)(t_4,t_1)}\left(\begin{array}{cc}
{d|D} & {h|H} \\
{c|C} & {b|B}\end{array}\right)
\mathbb{R}_{(t_4,t_1)(t_6,t_3)}\left(\begin{array}{cc}
{f|F} & {h|H} \\
{e|E} & {d|D}\end{array}\right),\qquad
\end{array}\qquad\qquad\qquad\qquad\qquad
\end{align}
where $(t_i,t_j)$ are the spectral parameters, and $\mathbb{R}$ is the face weight. We use the capital letters to define the discrete spin variables, and the small letters for continuous ones.

One can construct the solution to the Yang- Baxter equation from the solution of the star-triangle relation   \cite{Baxter:1997tn} by using the star-triangle relation several times. By using the solution (\ref{W})-(\ref{S}) one obtains the expression for the face weight  (see for more details \cite{Gahramanov:2015cva})
\begin{align} \nonumber
    \mathbb{R}_{(t_i,t_j)(t_k,t_l)}\left( \begin{array}{cc} {a|A} & {}{b|B}\\{f|F}&{}{h|H} \end{array}\right) & =\frac{(q^\frac23 (t/s)^{-2};q)_\infty( q^\frac23 (s/r)^{-2};q)_\infty(
q^\frac23 (r/t)^{-2};q)_\infty}{(q^\frac13 (t/s)^2;q)_\infty(
q^\frac13 (s/r)^2;q)_\infty(q^\frac13 (r/t)^2;q)_\infty} \\ \nonumber
  \qquad \,\,\, \,\,\,  \,\,\, \,\,\,\,\,\,\,\,\,\, \times \sum_{m}\int [d_{{m}}{z}]  & \mathcal{W}_{\frac{1}{6}+t_i-t_l}({a}|{A},{z}|{m}) \mathcal{W}_{\frac{1}{3}+t_j-t_i}({b}|{B},{z}|{m}) \\
& \times \mathcal{W}_{\frac{1}{3}+t_l-t_k}({f}|{F},{z}|{m})\mathcal{W}_{\frac{1}{6}+t_k-t_j}({h}|{H},{z}|{m}),       
\end{align}
where $\mathcal{W}$ is given in (\ref{W}), and the integral measure is defined as $[d_{{m}}{z}]:=q^{-m} (1-q^mz^2)(1-q^mz^{-2}) (4 \pi i z)^{-1} dz$.

\section{A new identity for the star-star relation }

In order to present a new multi-spin solution to  IRF-type Yang-Baxter equation, let us consider the following basic hypergeometric sum/integral which originates from three-dimensional supersymmetric index  computations:
\begin{align}
\mathcal{I}(\{t_j\},\{s_j\})= & \sum_{m_k} \int\prod_{k=1}^{n}\prod_{j=1}^{2n}\frac{\Big(q^{1+\frac{\tau_j+m_k}{2}}\frac{1}{t_jz_k};q\Big)_\infty }{\Big(q^{\frac{\tau_j+m_k}{2}}t_jz_k;q\Big)_\infty} 
\frac{\Big(q^{1+\frac{\kappa_j-m_k}{2}}\frac{z_k}{s_j};q\Big)_\infty }{\Big(q^{\frac{m_k-\kappa_j}{2}}\frac{s_j}{z_k};q\Big)_\infty}\\\nonumber &\qquad\qquad\qquad\qquad\qquad\times\frac{\Big(q^{\frac{m_k-m_j}{2}}z_j/z_k;q\Big)_\infty}{\Big(q^{1+\frac{m_j-m_k}{2}}z_k/z_j;q\Big)_\infty}
\prod_{k=1}^{n-1}\frac{dz_k}{z_k} .
\end{align}
This integral possesses a very interesting  transformation property 
\begin{align}\label{mainint}
\mathcal{I}(\{t_j\},\{s_j\})=\prod_{j,k=1}^{2n} \frac{\Big(q^{1+\frac{\tau_j+\kappa_k}{2}}\frac{1}{t_js_k};q\Big)_\infty }{\Big(q^{\frac{\tau_j+\kappa_k}{2}}t_js_k;q\Big)_\infty} \,\,\mathcal{I}(\{ \widetilde{t_j} \},\{\widetilde{s_j} \}),
\end{align}
with the balancing condition $\prod_{j=1}^{2n} s_j t_j = q^n$ and $\sum_{j=1}^{2n}\tau_j+\kappa_j=0$ and we used the following notations
 \begin{align}
\widetilde{t}_j=(\prod_{j=1}^{2n}t_j)^{\frac{1}{n}}t_j^{-1};\,\,\,\,\,
\widetilde{s}_j=(\prod_{j=1}^{2n}s_j)^{\frac{1}{n}}s_j^{-1},
\end{align}
and similarly for the discrete variables
\begin{align}
\widetilde{\tau}_j=\frac{\sum_{j=1}^{2n}\tau_j}{n}-\tau_j;\,\,\, \,\,\, \widetilde{\kappa}_j=\frac{\sum_{j=1}^{2n}\kappa_j}{n}-\kappa_j. 
\end{align}
The left-hand side of the integral identity (\ref{mainint}) gives the $V$--function introduced in \cite{Gahramanov:2015cva} when $n=2$
\begin{multline}\label{inindex}
{\bf
  V}(\{t_i\},\{s_i\}) =
 \sum_m\int [d_{{m}}{z}] \prod_{i=1}^4   \frac{(q^{1+(m+\tau_i)/2}/t_iz,q^{1+(\tau_i-m)/2} z/t_i ;q)_\infty}{(q^{(m+\tau_i)/2}t_iz,q^{(\tau_i-m)/2} t_i/z ;q)_\infty}  \\
 \times \frac{(q^{1+(m+\kappa_i)/2}/s_iz,q^{1+(\kappa_i-m)/2} z/s_i ;q)_\infty}{(q^{(m+\kappa_i)/2}s_iz,q^{(\kappa_i-m)/2} s_i/z ;q)_\infty}.
 \end{multline}
The integral identity (\ref{mainint}) is just $W(E_7)$ transformation property of the integral (\ref{inindex}) (see \cite{Gahramanov:2015cva,Spiridonov:2008zr,Gahramanov:2013xsa} for details).

The multi-spin generalization of the IRF-type Yang-Baxter equation has the following form
\begin{equation}
\begin{array}{l}
\displaystyle \sum_{\boldsymbol{H}}\int d\boldsymbol{h}\;
\mathbb{R}_{\boldsymbol{u|U}\,\,\boldsymbol{v|V}}\left(\begin{array}{cc}
\boldsymbol{c|C} & \boldsymbol{h|H} \\
\boldsymbol{e|E} & \boldsymbol{d|D}\end{array}\right)
\mathbb{R}_{\boldsymbol{u|U}\,\,\boldsymbol{w|W}}\left(\begin{array}{cc}
\boldsymbol{h|H} & \boldsymbol{b|B} \\
\boldsymbol{d|D} & \boldsymbol{f|F}\end{array}\right)
\mathbb{R}_{\boldsymbol{v|V}\,\,\boldsymbol{w|W}}\left(\begin{array}{cc}
\boldsymbol{c|C} & \boldsymbol{g|G} \\
\boldsymbol{h|H} & \boldsymbol{b|B}\end{array}\right)\\
[5mm]
\displaystyle \;\;\;\;\;\;\;\;\;\;\; =\sum_{\boldsymbol{H}}\int d\boldsymbol{h}\;
\mathbb{R}_{\boldsymbol{v|V}\,\,\boldsymbol{w|W}}\left(\begin{array}{cc}
\boldsymbol{e|E} & \boldsymbol{h|H} \\
\boldsymbol{d|D} & \boldsymbol{f|F}\end{array}\right)
\mathbb{R}_{\boldsymbol{u|U}\,\,\boldsymbol{w|W}}\left(\begin{array}{cc}
\boldsymbol{c|C} & \boldsymbol{g|G} \\
\boldsymbol{e|E} & \boldsymbol{h|H}\end{array}\right)
\mathbb{R}_{\boldsymbol{u|U}\,\,\boldsymbol{v|V}}\left(\begin{array}{cc}
\boldsymbol{g|G} & \boldsymbol{b|B} \\
\boldsymbol{h|H} & \boldsymbol{f|F}\end{array}\right),
\end{array}\label{YBE-IRF}
\end{equation}
where all bold letters are defined as $\boldsymbol{x}:= \{x_i\} \; \text{with} \,i=1,...,n$ for continuous spins, (and $\boldsymbol{A}:= \{A_i\} \; \; \text{with} \,i=1,...,n$ for discrete spins) and corresponds to the multi-spin components of the face weight. 

We repeat the arguments of \cite{Bazhanov:2011mz,Bazhanov:2013bh} in order to find the solution to IRF-type Yang-Baxter equation. Note that we just conjecture the solution, and do not carry out any detailed computations which is out of the context of this note. 

The multi-spin solution to the Yang-Baxter equation (\ref{YBE-IRF}) reads as
\begin{equation}
\mathbb{R}_{\boldsymbol{u|U}\boldsymbol{v|V}}\left(\begin{array}{cc}
  \boldsymbol{a|A} & \boldsymbol{b|B}\\ \boldsymbol{c|C} &
  \boldsymbol{d|D}\end{array}\right)\;=\;\varrho\; \left( \prod_{j,k=1}^{2n} \frac{\Big(q^{1+\frac{\tau_j+\kappa_k}{2}}\frac{1}{t_js_k};q\Big)_\infty }{\Big(q^{\frac{\tau_j+\kappa_k}{2}}t_js_k;q\Big)_\infty} \right)^{-1/2}\mathcal{I} (\{t_i\},\{s_i\})\;, \label{V1I}
\end{equation}
where we use the following notations in (\ref{V1I})
 \begin{align}
& t_j=q^{-2(u-v)-2 i c_j},\quad t_{n+j}
 =q^{-2(u'-v')-2 i b_j},\quad \\
 & \tau_j=-2(U-V)-2 i C_j,\quad \tau_{n+j}
 =-2(U'-V')-2 i B_j,\quad \\
& s_j=q^{2(u'-v-\eta)+2 i a_j}\;,\quad
 s_{n+j}=q^{2(u-v'-\eta)+2 i d_j}\;, \quad  \label{t-change} \\
 & \kappa_j=2(U'-V-\eta)+2 i A_j\;,\quad
 \kappa_{n+j}=2(U-V'-\eta)+2 i D_j\;, \quad \text{and $j=1,...,n$} \label{t-change} \\
 & \boldsymbol{u}=[u, u'], \quad \boldsymbol{v}=[v,v'] \;,\quad \boldsymbol{U}=[U, U'], \quad \boldsymbol{V}=[V,V'].
 \end{align}
Here 
\begin{equation}
\varrho=\frac{\sqrt{\mathbb{S}(\boldsymbol{c|C})\mathbb{S}(\boldsymbol{b|B})}}
{k_n(\eta-u+v)k_n(\eta-u'+v')k_n(u'-v)k_n(u-v')}\;.  
\label{varrho-def}  
\end{equation}
\begin{equation}
\label{Schange}
\mathbb{S}(\boldsymbol{x|X})= \frac{1}{2} \left( \prod_{j\not= k}
\frac{\Big(q^{1+\frac{X_j-X_k}{2}}\frac{z_k}{z_j};q\Big)_\infty }{\Big(q^{\frac{X_k-X_j}{2}}\frac{z_j}{z_k};q\Big)_\infty} \right)^{-1},\,\, \text{with}\,\, z_j=q^{2 i x_j},
\end{equation}
and $k_n$ is some normalizing parameter and for $n=2$ is given in (\ref{k}).

In terms of Boltzmann weights the expression (\ref{YBE-IRF}) gets the form of the star-star relation and reads as
\begin{align}\nonumber\label{St-st}
\mathcal{W}_{2-\gamma-\delta}(\boldsymbol{d|D,c|C}) \mathcal{W}_{2-\beta-\gamma}(\boldsymbol{b|B,c|C})\sum_{\boldsymbol{X}}\int [d_{\boldsymbol{X}}\boldsymbol{x}]\mathcal{S}(\boldsymbol{x|X})\mathcal{W}_{\alpha}(\boldsymbol{a|A,x|X})\\\nonumber
\qquad\times\mathcal{W}_{\beta}(\boldsymbol{x|X,b|B})\mathcal{W}_{\gamma}(\boldsymbol{c|C,x|X})\mathcal{W}_{\delta}(\boldsymbol{x|X,d|D})\qquad\qquad\\\nonumber
\qquad=\mathcal{W}_{2-\alpha-\beta}(\boldsymbol{a|A,b|B}) \mathcal{W}_{2- \gamma-\delta}(\boldsymbol{a|A,d|D})\sum_{\boldsymbol{X}}\int [d_{\boldsymbol{X}}\boldsymbol{x}]\mathcal{S}(\boldsymbol{x|X})\mathcal{W}_{\gamma}(\boldsymbol{x|X,a|A})\\\times
\mathcal{W}_{\delta}(\boldsymbol{b|B,x|X})\mathcal{W}_{\alpha}(\boldsymbol{x|X,c|C})\mathcal{W}_{\beta}(\boldsymbol{d|D,x|X}).
\end{align}
where $\alpha+\beta+\gamma+\delta=2$, and the integral measure is
\begin{align}
  \sum_{\boldsymbol{X}}\int [d_{\boldsymbol{X}}{\boldsymbol{x}}] := \sum_{X_1,X_2,..X_n}\int \frac{dx_1}{2\pi i x_1} \frac{dx_2}{2\pi i x_2}\cdots\frac{dx_n}{2\pi i x_n}.
\end{align}

Note that one can obtain this solution by taking certain limit from the solution presented in \cite{Yamazaki:2013nra}. By taking limit of (\ref{V1I}) one may construct new solution to the Yang-Baxter equation. For instance, using the following property of the q-Pochhammer symbol
\begin{equation}
        \lim_{q \rightarrow 1} \frac {(q^\alpha;q)_\infty}{(q^\beta;q)_\infty} (1-q)^{\alpha-\beta}= \frac{ \Gamma (\beta)}{ \Gamma (\alpha)},
\end{equation}
one obtains the solution in terms of Euler's gamma function\footnote{Special form of this solution ($n=2$ case)  is presented in \cite{Kels:2013ola}.} \cite{Yamazaki:2016wnu}
\begin{align}
\mathcal{W}_\alpha(\boldsymbol{x|X,y|Y})=\prod_{i,j=1}^n\frac{\Gamma(\alpha-\eta \pm[\frac{(X_i+Y_j)}{2}+ i(x_i+y_j)])}{\Gamma(1+\eta-\alpha \pm [-\frac{(X_i+Y_j)}{2} + i(x_i+y_j)])} ,
\end{align}
\begin{align}
\mathcal{S}(\boldsymbol{x}|\boldsymbol{X})=\frac{1}{n!}\prod_{i\neq j}\Big((x_i-x_j)^2+(\frac{X_i-X_j}{2})^2\Big).
\end{align}

\section{Three-dimensional supersymmetric index}
In this section, we give a brief information about three dimensional supersymmetric index which we use to get the solution to the Yang-Baxter equation.  

In order to obtain the star-star relation we use the so-called gauge/YBE correspondence. The essence of the correspondence may be explained in simple terms as follows. The equality of partition functions of supersymmetric dual theories may be written as the Yang-Baxter equation. Therefore the exact computation of the partition function on supersymmetric gauge theory side leads to an  integrable lattice model.

The supersymmetric index of three--dimensional ${\mathcal N}=2$ supersymmetric field theory is a twisted partition function defined on $S^2 \times S^1$:
\begin{equation}
\mathcal{I}(q,t)=\text{Tr} \left[ (-1)^\text{F} e^{-\beta \{Q, Q^\dagger \} }
q^{\frac12(\Delta+j_3)}\prod_i t_i^{F_i} \right],
\end{equation}
where $F$ is the fermion number, $Q$ and $Q^\dagger$ are corresponding supercharges,  $\Delta$ is the energy, $j_3$ is the third component of the angular momentum around $S^2$, and $F_i$'s are the Cartan generators of the flavor symmetry and $t_i$'s are corresponding fugacities, and the trace is taken over the Hilbert space of the theory. Using the localization technique \cite{Pestun:2007rz} the supersymmetric index can be computed exactly, and it gets the form of the matrix integral  \cite{Kim:2009wb,Imamura:2011su}.

The identity (\ref{mainint}) in terms of supersymmetric gauge theory computations corresponds to  the equality of supersymmetric indices of the following Seiberg dual theories:

\noindent - \textbf{the left-hand side} of the identity  is the index of three-dimensional $\mathcal N=2$ Supersymmetric Quantum Chromodynamics with gauge group $SU(N)$ and flavor group $SU(2N) \times SU(2N)$, chiral multiplets belonging to the fundamental and anti-fundamental representations of gauge group and corresponding flavor group; 

\noindent - \textbf{the right-hand side} of the identity  is the index of the dual theory with $SU(N)$ gauge group and with the same flavor group, including mesons in the fundamental and anti-fundamental representation of the flavor group.

\section*{Acknowledgements}
We would like to thank Nesin Mathematical Village and organizers of the Fifth Autumn School and Workshop of RTN where some parts of the work was done for the warm hospitality. Sh.J is also grateful to Max-Planck-Institut f\"{u}r Gravitationsphysik (Albert-Einstein-Institut) for the hospitality. Special thanks are to Cestmir Burdik and to other organizers of the XXVth International Conference on Integrable Systems and Quantum symmetries (ISQS-25) for the invitation and support.\\

\medskip

\bibliographystyle{utphys}
\bibliography{YBE}

\end{document}